\documentclass[lettersize,journal]{IEEEtran}
\usepackage{amsmath,amsfonts}
\usepackage{amssymb}
\usepackage{amsthm}
\usepackage{algorithm}
\usepackage{algorithmic}
\usepackage{array}
\usepackage{textcomp}
\usepackage{stfloats}
\usepackage{url}
\usepackage{verbatim}
\usepackage{graphicx}
\usepackage{bm}
\usepackage[normalem]{ulem}
\usepackage[tableposition=top]{caption}
\usepackage{subcaption}

\newtheoremstyle{italic}
  {\parskip}       
  {\parskip}       
  {\normalfont}    
  {\parindent}     
  {\itshape}       
  {:}              
  {.5em}           
  {}               
\theoremstyle{italic}

\usepackage[capitalize]{cleveref}
\usepackage{cite}
\usepackage{color}
\usepackage{booktabs}
\usepackage{lettrine}
\begin{document}

\title{SA-RA-JSCC: SNR-Adaptive and Semantic-Rate-Aware Joint Source-Channel Coding}


\author{~Shitong~Zhang,~Yaping~Sun,~\IEEEmembership{Member,~IEEE},~Hao Chen,~\IEEEmembership{Member,~IEEE},~Xiaoyi~Li,\\~Bo~Gu,~\IEEEmembership{Member,~IEEE},~Xiaodong~Xu,~\IEEEmembership{Senior Member,~IEEE}, and Nan~Ma,~\IEEEmembership{Member,~IEEE}

\thanks{Shitong Zhang is with Department of Broadband Communication, Pengcheng Laboratory, Shenzhen 518066, China, and also with the School of Intelligent Systems, Sun Yat-sen University, Shenzhen 518107, China (e-mail:~zhangsht35@mail2.sysu.edu.cn).

Yaping Sun is with Department of Broadband Communication, Pengcheng Laboratory, Shenzhen 518066, China (e-mail:~sunyp@pcl.ac.cn).

Hao Chen is with Department of Broadband Communication, Pengcheng Laboratory, Shenzhen 518066, China (e-mail:~chenh03@pcl.ac.cn).

Xiaoyi li and Bo Gu are with the School of Intelligent Systems, Sun Yat-sen University, Shenzhen 518107, China, also with Southern Marine Science and Engineering Guangdong Laboratory (Zhuhai), and also with Guangdong Provincial Key Laboratory of Fire Science and Technology, Guangzhou 510006, China (e-mail:~lixy627@mail2.sysu.edu.cn,~gubo@mail.sysu.edu.cn).

Xiaodong Xu and Nan Ma are with Beijing University of Posts and Telecommunications, Beijing 100876, China, and with Department of Broadband Communication, Pengcheng Laboratory, Shenzhen 518066, China (e-mail:~xuxiaodong@bupt.edu.cn,~manan@bupt.edu.cn).
}

}
\maketitle

\begin{abstract}
In joint source-channel coding (JSCC)-based semantic communication systems, achieving stable and reliable image semantic transmission under channel constraints remains a key challenge. In most channel adaptation modules, the signal-to-noise ratio (SNR) is often injected into each layer of a channel-adaptation model in an independent and layer-wise manner, which undermines global coordination across layers. Therefore, consistent noise-robust representations may fail to be learned throughout the model. To address this problem, we propose SA-RA-JSCC, a novel channel-adaptive JSCC model. SA-RA-JSCC maps SNR into a unified semantic vector in the feature space and then applies a one-shot global reweighting to the encoded features, thereby enabling globally consistent and learnable channel adaptation. Moreover, in order to further enhance the anti-channel capability of semantic information, a semantic-rate-aware module is introduced, enabling the adaptive policy to respond simultaneously to fluctuations in channel quality and changes in semantic-rate constraints, thereby enhancing global network coordination and channel adaptivity. Extensive experiment results across multiple channels and datasets demonstrate that SA-RA-JSCC significantly outperforms existing semantic communication models in terms of reconstruction metrics such as PSNR and MS-SSIM, exhibiting stronger robustness across a broad range of SNR regimes.
\end{abstract}


\section{Introduction}
Semantic communication has emerged as a salient candidate for the next generation of wireless systems. Leveraging deep learning (DL)-enabled joint source–channel coding (JSCC), this paradigm dispenses with the conventional separation principle and instead learns, in an end-to-end fashion, to convey the meaning embedded in the source data. In doing so, it holds the promise of simultaneously improving reliability and transmission efficiency relative to classical digital architectures. As an influential early contribution, DeepJSCC \cite{8723589} introduced a convolutional neural network (CNN)-based end-to-end design that directly maps images to channel symbols, delivering superior reconstruction performance—most notably in low signal-to-noise ratio (SNR) regimes. However, CNN encoders are constrained by local receptive fields and largely uniform compression, which tend to treat heterogeneous regions indiscriminately and thus limit the model's capacity to represent and preserve semantically pivotal content. To address these limitations, subsequent works have turned to Transformer-based semantic encoders \cite{9653664, 10833860, 10589474}. Unlike fixed-kernel convolutions that primarily capture local patterns, the self-attention mechanism in Transformers can model long-range dependencies at the global level and adaptively aggregate multi-scale semantic cues, allowing the encoder to allocate representational capacity toward task-relevant structures. Consequently, Transformer-based approaches have been shown to consistently outperform their CNN-based counterparts across a range of settings.

To enable robust semantic transmission under the JSCC paradigm, semantic communication commonly incorporates channel adaptation (CA) mechanisms to attenuate the perturbations induced by time-varying channel conditions on the end-to-end mapping. In this context, SwinJSCC \cite{10589474} is among the first to introduce SNR-adaptive (SA) and semantic-rate-adaptive (RA) modules at the encoder, reshaping semantic representations in a noise-aware manner prior to channel passage and thereby substantially improving reconstruction fidelity. However, by injecting SNR and semantic-rate independently at each layer, this design lacks a cross-layer, globally coherent adaptation mechanism, making it prone to inducing mutually incompatible denoising behaviors across depths. Such inconsistency can precipitate representational conflicts among hierarchical features and ultimately erode overall robustness. \cite{10500305} leverages channel-output feedback to adaptively modulate the encoder outputs, yet this strategy hinges on an additional feedback link and thus incurs nontrivial communication overhead. By contrast, \cite{10015684, 9878262} introduce a channel-attention mechanism during compression, embedding SNR-related information into each attention layer to better exploit channel-state cues. While these approaches enhance channel awareness and adaptivity to some extent, their gains are frequently purchased at the cost of substantial parameter storage and computational burden, rendering it difficult to sustain scalable efficiency and stable performance in practical deployments.

Taking inspiration from the above analysis, we propose SA-RA-JSCC, a novel channel-adaptive JSCC model built upon the SwinJSCC semantic encoding framework. Instead of injecting the SNR at each layer, SA-RA-JSCC maps them into a unified semantic vector in the feature space and then performs a one-shot global reweighting of the encoded features, yielding globally coordinated and learnable channel adaptation without altering the backbone architecture. Moreover, a semantic-rate module is incorporated and jointly leveraged with SNR in the adaptive policy, allowing the model to respond simultaneously to time-varying channel conditions and semantic-rate constraints. Extensive experiments across multiple compression rate and datasets verify that SA-RA-JSCC consistently improves reconstruction performance in terms of PSNR and MS-SSIM, while exhibiting stronger robustness across a broad range of SNR regimes and improved cross-dataset generalization.

The rest of this letter is organized as follows. The system model will be introduced in the Section~\ref{sec:sm}. The Section~\ref{sec:as} explains the SA-RA-JSCC details. The SECTION~\ref{sec:sa} conducts experiment analysis. The SECTION~\ref{sec:c} provides a conclusion of the entire letter.
\section{System Model}\label{sec:sm}
\begin{figure*}[htpb]
  \centering
  \subfloat[The architecture of the SA-RA-JSCC.]{\includegraphics[width=8.5cm,trim=2pt 2pt 2pt 2pt,clip]{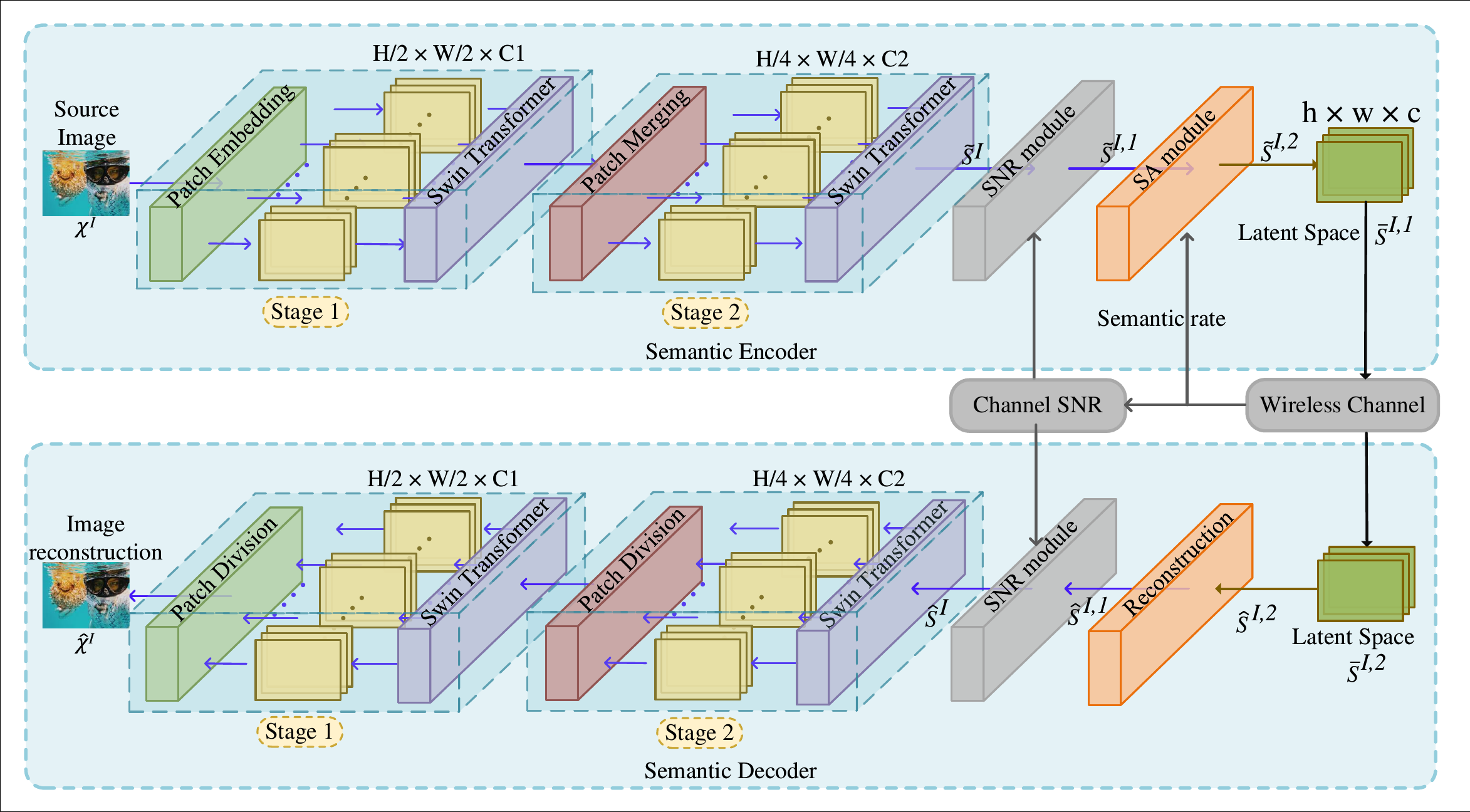}\label{fig:Kodak}}\hspace{1em}
  \subfloat[The architecture of SA module and RA module.]{\includegraphics[width=8.5cm,trim=2pt 2pt 2pt 2pt,clip]{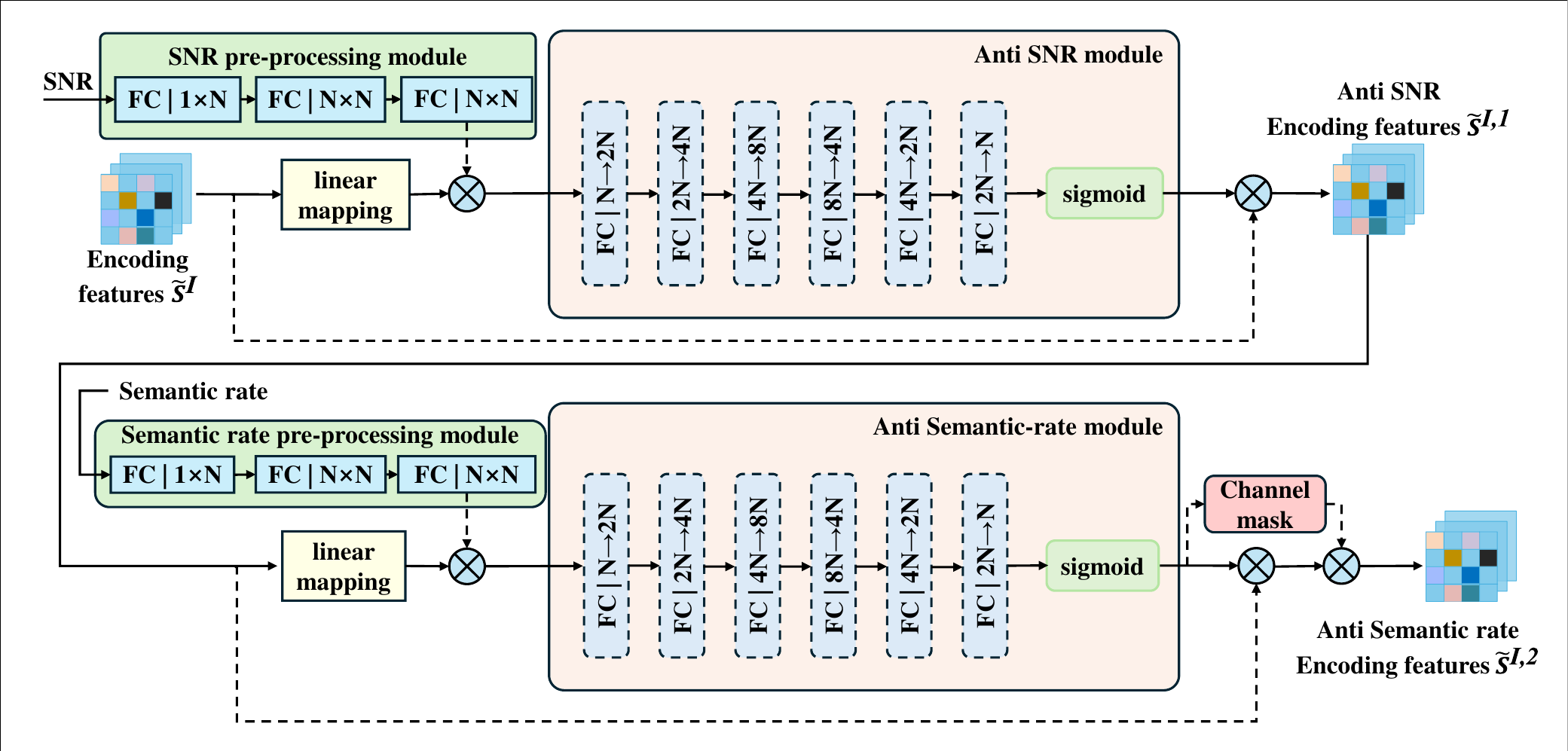}\label{fig:CLIC}}
  \caption{Illustration of the proposed SA-RA-JSCC.}
  \label{Fig:1}
\end{figure*}
Fig.~\ref{Fig:1}(a) illustrates an overview of the designed SA-RA-JSCC wireless image transmission architecture. The input RGB image $\bm{\chi} \in \mathbb{R}^{H \times W \times 3}$ is first fed into the patch embedding layer of the encoder. In stage~1, the image is partitioned into $\frac{H}{2}\times\frac{W}{2}$ non-overlapping patches, which are embedded into token features with channel dimension $C_1$. Accordingly, the stage-1 feature map is represented as $\mathbb{R}^{\frac{H}{2}\times\frac{W}{2}\times C_1}$, where $H$ and $W$ denote the height and width of $\bm{\chi}$. These tokens are then processed by $n_1$ Swin Transformers \cite{10589474}. The Swin Transformers integrate standard multi-head self-attention (MSA) modules and feed-forward networks to process the tokens from the previous layer. The shifted window-based self-attention mechanism enables the model to capture long-range dependencies within the image. Afterward, the tokens from stage 1 are merged through patch merging in stage 2, resulting in patches of size $g_2\in\mathbb{R}^{\frac{H}{4}\times\frac{W}{4}\times C_2}$, where $C_2$ is the channel dimension of stage 2. These merged patches are then processed by $n_2$ Swin Transformers for further learning. This process is performed in two stages. By this method, the proposed model significantly enhances the model capacity, enabling the efficient learning of complex details in high-resolution images by leveraging global information. The learned semantic features are then fed into the channel adaptation module, incorporating both SNR adaptation and semantic-rate adaptation. This ensures that the semantic features exhibit improved noise robustness. The final semantic feature vector $\overline{\bm{S}}^{I,1}\in\mathbb{R}^{h\times w\times c}$ represents the semantic latent representation of the input image $\bm{\chi}$, where $h$, $w$ and $c$ are the dimensions that enter the latent semantic space, respectively.  

For the wireless channel, we adopt a general fading channel model characterized by the transfer function $\overline{\bm{S}}^{I,2} = \bm{h}_c \bigodot \overline{\bm{S}}^{I,1} + \bm{n}_c$. Here, $\bm{h}_c$ denotes the channel state information (CSI) vector, and $\bm{n}_c$ represents the additive noise vector. The components of the noise vector are mutually independent and follow a Gaussian distribution, as denoted $\bm{n}_c \sim \mathcal{N}(0, \sigma_{\bm{n}_c}^2)$, where $\sigma_{\bm{n}_c}$ denotes the average noise power.

The decoder adopts a symmetric architecture similar to the encoder, consisting of feature reconstruction, channel module, patch division for upsampling, and Swin Transformers, forming the complete SA-RA-JSCC encoder-decoder framework.

\section{SA-RA-JSCC}\label{sec:as}
\subsection{SNR-Adaptive Module}
In this paper, a novel SNR-adaptive module is proposed, which significantly improves image reconstruction performance even under severely degraded channel conditions. The proposed architecture of SNR-adaptive module is shown top in Fig.~\ref{Fig:1}(b). The SNR pre-processing module is defined as follows:
\begin{align}\label{eq:1}
\left\{\begin{aligned}
\bm{S}_{mi}^{(1)}=&\mathrm{max}\left(0, \bm{W}_{mi}^{(1)}\cdot\gamma_I+\bm{b}_{mi}^{(1)}\right),\\
\bm{S}_{mi}^{(2)}=&\mathrm{max}\left(0, \bm{W}_{mi}^{(2)}\cdot \bm{S}_{mi}^{(1)}+\bm{b}_{mi}^{(2)}\right),\\
\bm{S}_{mi}=&\mathrm{Sigmoid}\left(\bm{W}_{mi}^{(3)}\cdot \bm{S}_{mi}^{(2)}+\bm{b}_{mi}^{(3)}\right),
\end{aligned}\right.
\end{align}
where $\bm{W}_{mi}^{(n)}\in\mathbb{R}^{N\times N},n\in\{1,2,3\}$ and $\bm{b}_{mi}^{(n)}\in\mathbb{R}^{N\times 1},n\in\{1,2,3\}$ respectively denote the weights and biases of neural networks (NNs). $\gamma_I$ is the SNR.

Assuming the encoded semantic feature vector is $\tilde{\bm{S}}^I$, the anti-SNR encoding features $\tilde{\bm{S}}^{I, 1}$ after SNR-adaptive module is represented as:
\begin{align}\label{eq:2}
\begin{aligned}
\tilde{\bm{S}}^{I, 1}=\tilde{\bm{S}}^I + \mathrm{Sigmoid}\left(\mathrm{MLP}(\bm{S}_{mi}\cdot \tilde{\bm{S}}^I)\right),
\end{aligned}
\end{align}
where $\mathrm{Sigmoid}$ is the activation function and $\mathrm{MLP}$ is the multilayer perceptron. Through multi-layer learning, the characteristics of SNR can be better learned.
\subsection{Semantic-Rate-Aware Module} 
In order to better perceive the features of semantic-rate, we have also designed a new semantic-rate-aware module, as shown below in Fig.~\ref{Fig:1}(b). The semantic-rate pre-processing module is defined as follows:
\begin{align}\label{eq:3}
\left\{\begin{aligned}
\bm{S}_{mj}^{(1)}=&\mathrm{max}\left(0, \bm{W}_{mj}^{(1)}\cdot R_I+\bm{b}_{mj}^{(1)}\right),\\
\bm{S}_{mj}^{(2)}=&\mathrm{max}\left(0, \bm{W}_{mj}^{(2)}\cdot \bm{S}_{mj}^{(1)}+\bm{b}_{mj}^{(2)}\right),\\
\bm{S}_{mj}=&\mathrm{Sigmoid}\left(\bm{W}_{mj}^{(3)}\cdot \bm{S}_{mj}^{(2)}+\bm{b}_{mj}^{(3)}\right),
\end{aligned}\right.
\end{align}
where $\bm{W}_{mj}^{(n)}\in\mathbb{R}^{N\times N},n\in\{1,2,3\}$ and $\bm{b}_{mj}^{(n)}\in\mathbb{R}^{N\times 1},n\in\{1,2,3\}$ respectively denote the weights and biases of NNs, in the same way as the SNR-Adaptive Module. $R_I$ is the actual semantic-rate. We already know that anti-SNR encoding features is $\tilde{\bm{S}}^{I, 1}$ after passing through the SNR-adaptive module. Therefore, the anti-semantic-rate encoding features before channel mask can be designed as:
\begin{align}\label{eq:4}
\begin{aligned}
\hat{\bm{S}}^{I, 2}=\tilde{\bm{S}}^{I,1} + \mathrm{Sigmoid}\left(\mathrm{MLP}(\bm{S}_{mj}\cdot \tilde{\bm{S}}^{I, 1})\right).
\end{aligned}
\end{align}

In addition, the model framework in semantic-rate-aware module is similar to that in adaptive SNR-aware module. The anti-SNR module and anti-semantic-rate module share the same structure. However, relying solely on the learned module $\hat{\bm{S}}^{I, 2}$ to adapt to the semantic-rate is insufficient. Therefore, following \cite{10589474}, we introduce a code-mask module to analyze the relevance of semantic features to the semantic-rate and rank them along the channel dimensions. We then select the top $C$ dimensions from the ranking to construct a binary mask vector $\bm{p}$, where the first $C$ entries are set to 1 and the remaining entries are 0. The anti-semantic-rate encoding features are represented as:
\begin{align}\label{eq:5}
\begin{aligned}
\tilde{\bm{S}}^{I, 2}=\langle \hat{\bm{S}}^{I,2} \cdot \bm{p}\rangle.
\end{aligned}
\end{align}
This strategy enhances the proposed model's adaptability to semantic-rate variations by reweighting the semantic features that are most relevant to the target rate.
\subsection{Loss Function of SA-RA-JSCC}
Due to the varying resolutions of the input images, we train the proposed model using the widely adopted perceptual metric MS-SSIM and pixel-wise metric PSNR. Specifically, for MS-SSIM, the loss function is defined as:
\begin{align}\label{eq:6}
\begin{aligned}
\mathcal{L_M} = 1 - \big[l_M(\bm{\chi},\hat{\bm{\chi}})\big]^{\alpha_M}
\prod_{j=1}^{M}\big[c_j(\bm{\chi},\hat{\bm{\chi}})\big]^{\beta_j}
\big[s_j(\bm{\chi},\hat{\bm{\chi}})\big]^{\gamma_j},
\end{aligned}
\end{align}
where $M$ is the MS-SSIM scale number. $\alpha_M$, $\beta_j$, and $\gamma_j$ are the weight coefficients for each scale. At the $j$-th scale, the local statistics are the means $\mu_{\bm{\chi}}$ and $\mu_{\hat{\bm{\chi}}}$, the standard deviations $\sigma_{\bm{\chi}}$ and $\sigma_{\hat{\bm{\chi}}}$, and the covariance $\sigma_{{\bm{\chi}}\hat{\bm{\chi}}}$. The three components at a single scale are given by:
\begin{align}\label{eq:7}
\left\{\begin{aligned}
&l_M(\bm{\chi},\hat{\bm{\chi}})=\frac{2\mu_{\bm{\chi}}\mu_{\hat{\bm{\chi}}}+Z_1}{\mu_{\bm{\chi}}^2+\mu_{\hat{\bm{\chi}}}^2+Z_1},\\
&c_j(\bm{\chi},\hat{\bm{\chi}})=\frac{2\sigma_{\bm{\chi}}\sigma_{\hat{\bm{\chi}}}+Z_2}{\sigma_{\bm{\chi}}^2+\sigma_{\hat{\bm{\chi}}}^2+Z_2},\\
&s_j(\bm{\chi},\hat{\bm{\chi}})=\frac{\sigma_{\bm{\chi}\hat{\bm{\chi}}}+Z_3}{\sigma_{\bm{\chi}}\sigma_{\hat{\bm{\chi}}}+Z_3},
\end{aligned}\right.
\end{align}
where $Z_1$, $Z_2$, and $Z_3$ are the constant terms. For PSNR, we optimize the model by minimizing the mean squared error (MSE) between the source image $\bm{\chi}$ and the reconstructed image $\hat{\bm{\chi}}$.

\section{Experiment Analysis}\label{sec:sa}
\subsection{Experiment Setting}
The proposed scheme is evaluated on the widely used DIV2K dataset~\cite{agustsson2017ntire}. This dataset contains 1{,}000 RGB images with resolutions of approximately 2K, of which 800, 100, and 100 images are used for training, validation, and testing, respectively. During training, all images are randomly cropped into 256 $\times$ 256 patches to maintain consistent input dimensions. The batch size is set to 8. In addition, in the model of SA-RA-JSCC, we set the depth of 4 layers, where Swin Transformer blocks of each layer's depth is [2, 2, 6, 2], and the sizes of channel dimension $C$ are $[C_1, C_2, C_3, C_4]$ = [128, 192, 256, 320]. The size of the moving window for Swin Transformer is set to 8. The learning rate of the Adam optimizer is set to 0.0001. All experiments were trained and tested on NVIDIA GeForce RTX 5060 Ti.

To demonstrate the superiority of the proposed SA-RA-JSCC method, we compare it with four baselines:
\begin{enumerate}
  \item \textbf{SNR-EQ-JSCC \cite{10833860}}: This method embeds the SNR into the attention blocks and dynamically adjusts attention scores via channel embedding and query, yielding a channel-adaptive JSCC model.
  \item \textbf{SwinJSCC w/o CA \cite{10589474}}: SwinJSCC model without added SNR and semantic-rate modules.
  \item \textbf{SwinJSCC w/ CA \cite{10589474}}: SwinJSCC model with added SNR and semantic-rate modules.
  \item \textbf{ADJSCC \cite{9438648}}: This method adopts a CNN as the backbone and introduces channel attention to adaptively reweight feature channels, enabling channel-adaptive transmission.
  \item \textbf{BPG+LDPC \cite{bellard2018bpg,8316763}}: This scheme adopts the conventional separate source and channel coding framework, where the BPG \cite{bellard2018bpg} is used for compression and the 5G LDPC \cite{8316763} is employed for channel coding. Here, 5G LDPC codes with a block length of 8448 and 3840 bits are considered under different coding rates and quadrature amplitude modulation (QAM) schemes.
\end{enumerate}

In addition, to comprehensively evaluate the reconstruction quality of the proposed scheme, we adopt both the pixel-level metric PSNR and the perceptual metric MS-SSIM. The remaining parameters follow the SwinJSCC settings. Following \cite{10833860}, the compression rate $r$ is defined as follows:
\begin{align}\label{eq:8}
\begin{aligned}
r=\frac{h\times w \times c}{H \times W \times 3}.
\end{aligned}
\end{align}
Following \cite{10589474}, the channel bandwidth ratio $cbr$ is defined as:
\begin{align}\label{eq:9}
\begin{aligned}
cbr=\frac{c}{2\times 3 \times 2^i \times 2^i},
\end{aligned}
\end{align}
where $i$ is the number of the stages.
\subsection{Experiment Comparison Results of Different JSCC Schemes}
\begin{figure*}[htpb]
  \centering
  \subfloat[$r$ = 1/32]{\includegraphics[width=0.23\textwidth]{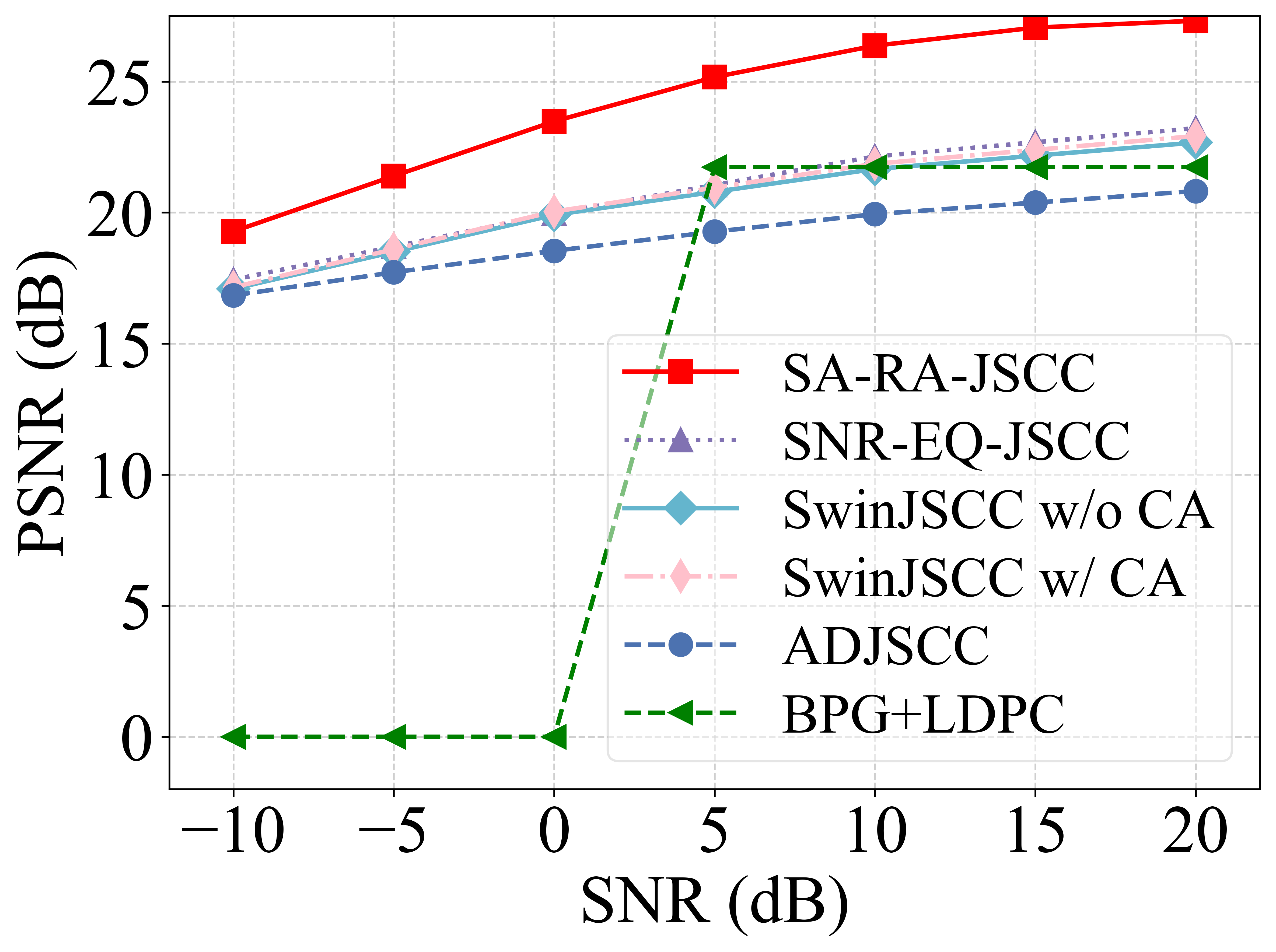}\label{fig:Kodak}}\hspace{0.5em}
  \subfloat[$r$ = 1/32]{\includegraphics[width=0.23\textwidth]{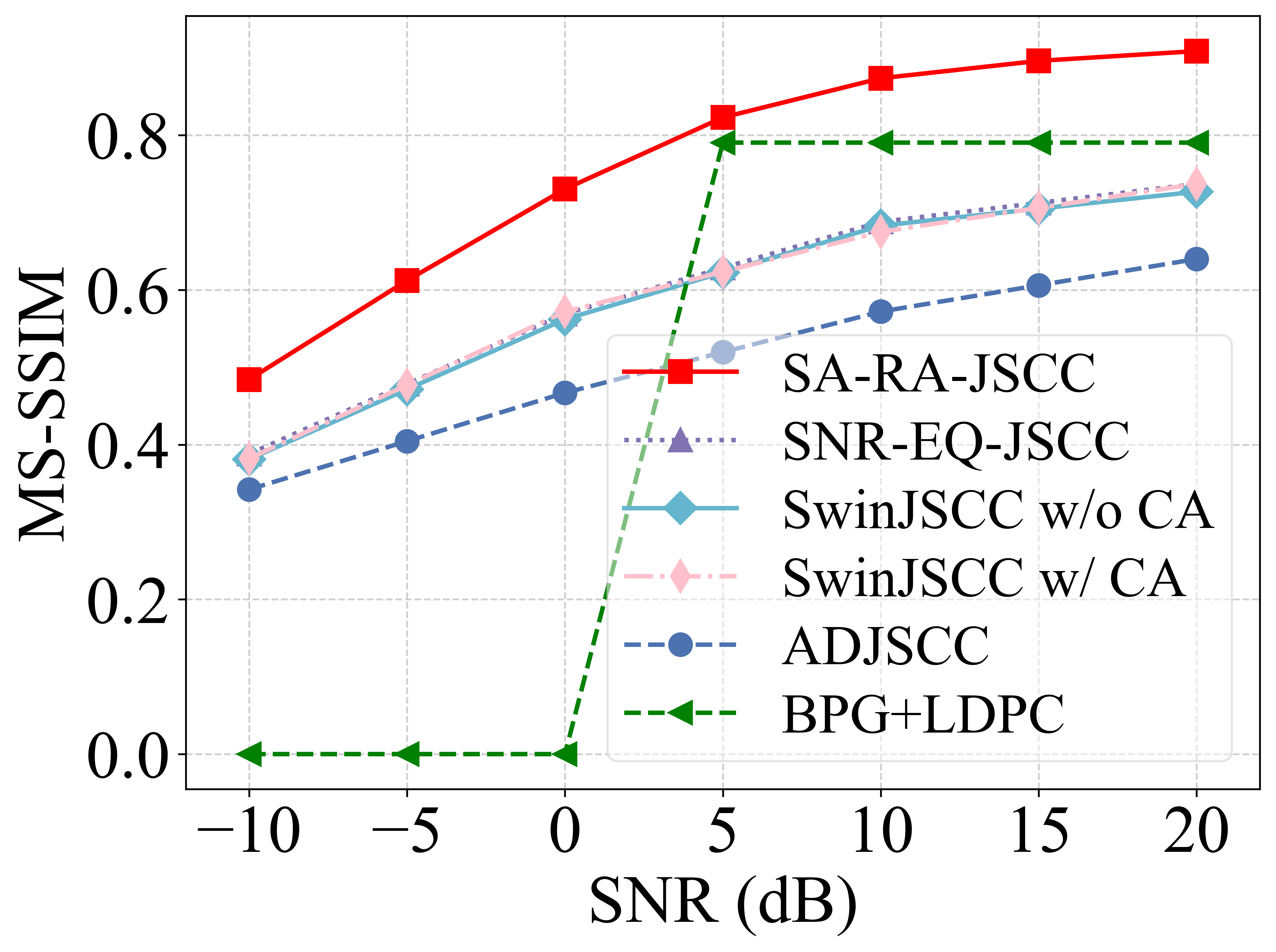}\label{fig:CLIC}}\hspace{0.5em}
  \subfloat[$r$ = 1/8]{\includegraphics[width=0.23\textwidth]{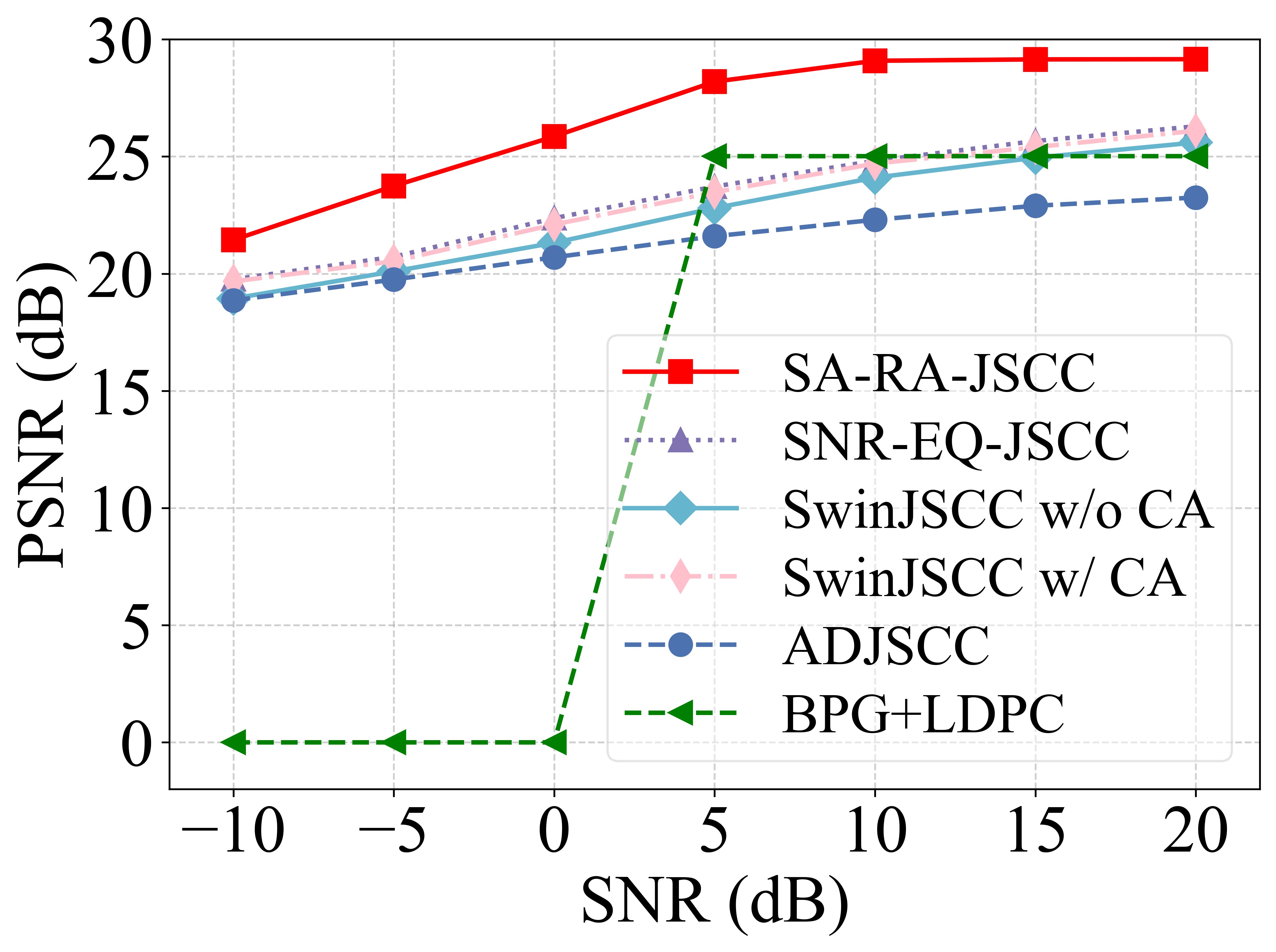}\label{fig:CLIC}}\hspace{0.5em}
  \subfloat[$r$ = 1/8]{\includegraphics[width=0.23\textwidth]{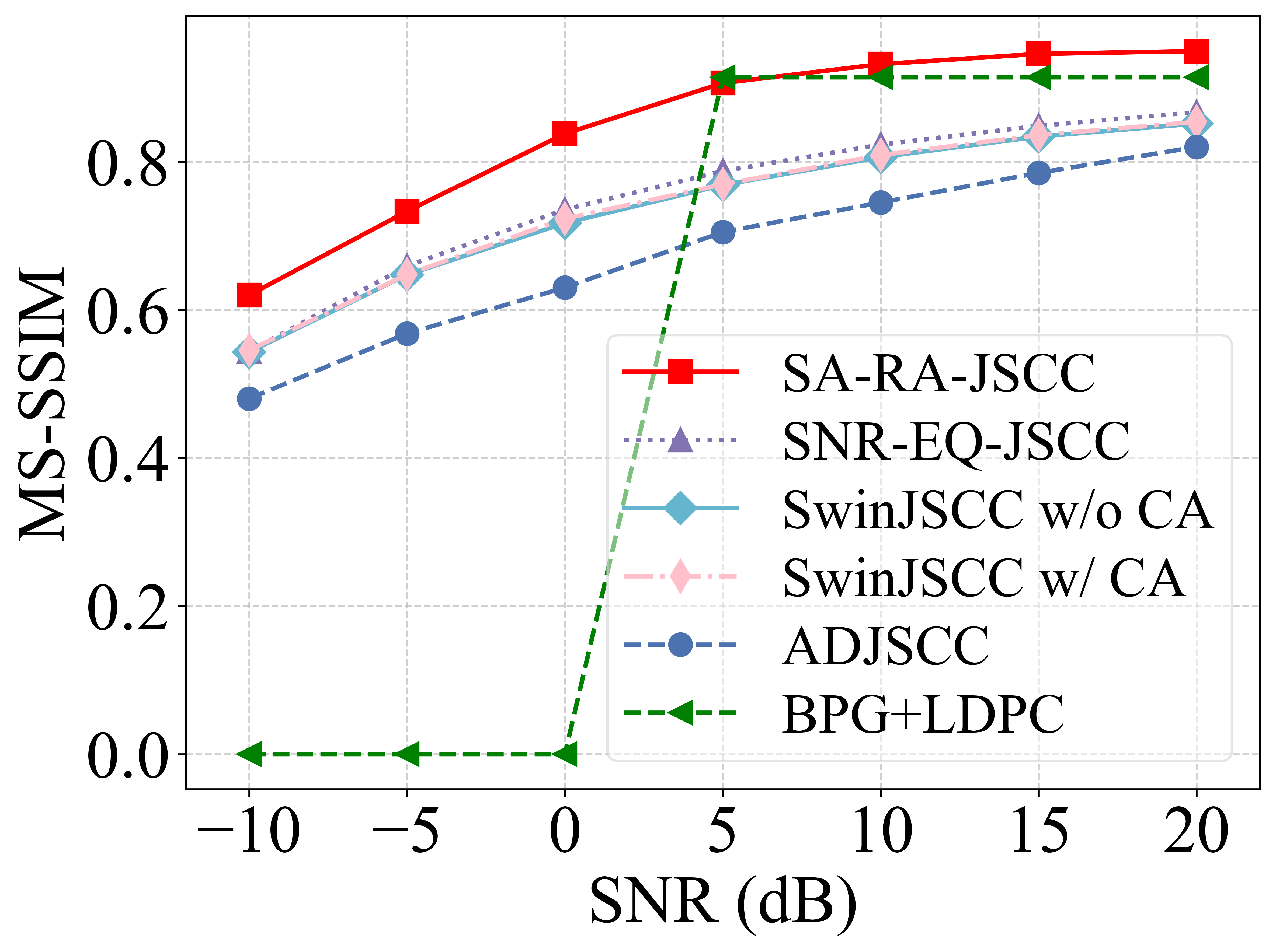}\label{fig:CLIC}}
  \caption{Comparison of different methods across various SNRs on the DIV2K dataset.}
  \label{Fig:3}
\end{figure*}

Fig.~\ref{Fig:3}(a) and Fig.~\ref{Fig:3}(b) compare PSNR and MS-SSIM versus SNR under the more stringent $r$ = 1/32, while Fig.~\ref{Fig:3}(c) and Fig.~\ref{Fig:3}(d) display the results under less stringent $r$ = 1/8 in the additive white Gaussian noise (AWGN) channel. Under $r$ = 1/32, where the latent representation is highly compressed, SA-RA-JSCC keeps a clear margin over all baselines at every SNR point. The improvement indicates that the proposed design is more effective at preserving and transmitting the most informative semantic features. In particular, its advantage over SNR-EQ-JSCC and SwinJSCC w/ CA suggests that adapting only to SNR is not sufficient under severe compression, whereas jointly handling channel condition and compression allows better feature prioritization and suppression of redundancy. Compared with the BPG+LDPC scheme, SA-RA-JSCC also exhibits significantly stronger robustness in the low- and medium-SNR regions. Since BPG+LDPC follows a separate source-channel coding paradigm and is sensitive to channel impairments, its reconstruction quality degrades sharply when the channel condition is poor. In contrast, the proposed joint design avoids the abrupt degradation effect and provides much more gradual performance degradation. The same trend is reflected in MS-SSIM, where SA-RA-JSCC maintains the best structural similarity throughout the sweep. When $r$ = 1/8, all methods improve, but SA-RA-JSCC remains consistently ahead and the gap becomes more visible at medium-to-high SNRs. This behavior implies that SA-RA-JSCC not only withstands harsh compression, but also makes better use of improved channel conditions to translate SNR gains into higher-fidelity reconstructions.  Although the BPG+LDPC baseline becomes competitive at high SNRs, it still remains inferior to SA-RA-JSCC, which demonstrates the advantage of semantic-aware joint source-channel optimization over the traditional separated coding architecture.

\begin{figure}[htpb]
  \centering
  \subfloat[CLIC2020]{\includegraphics[width=0.23\textwidth]{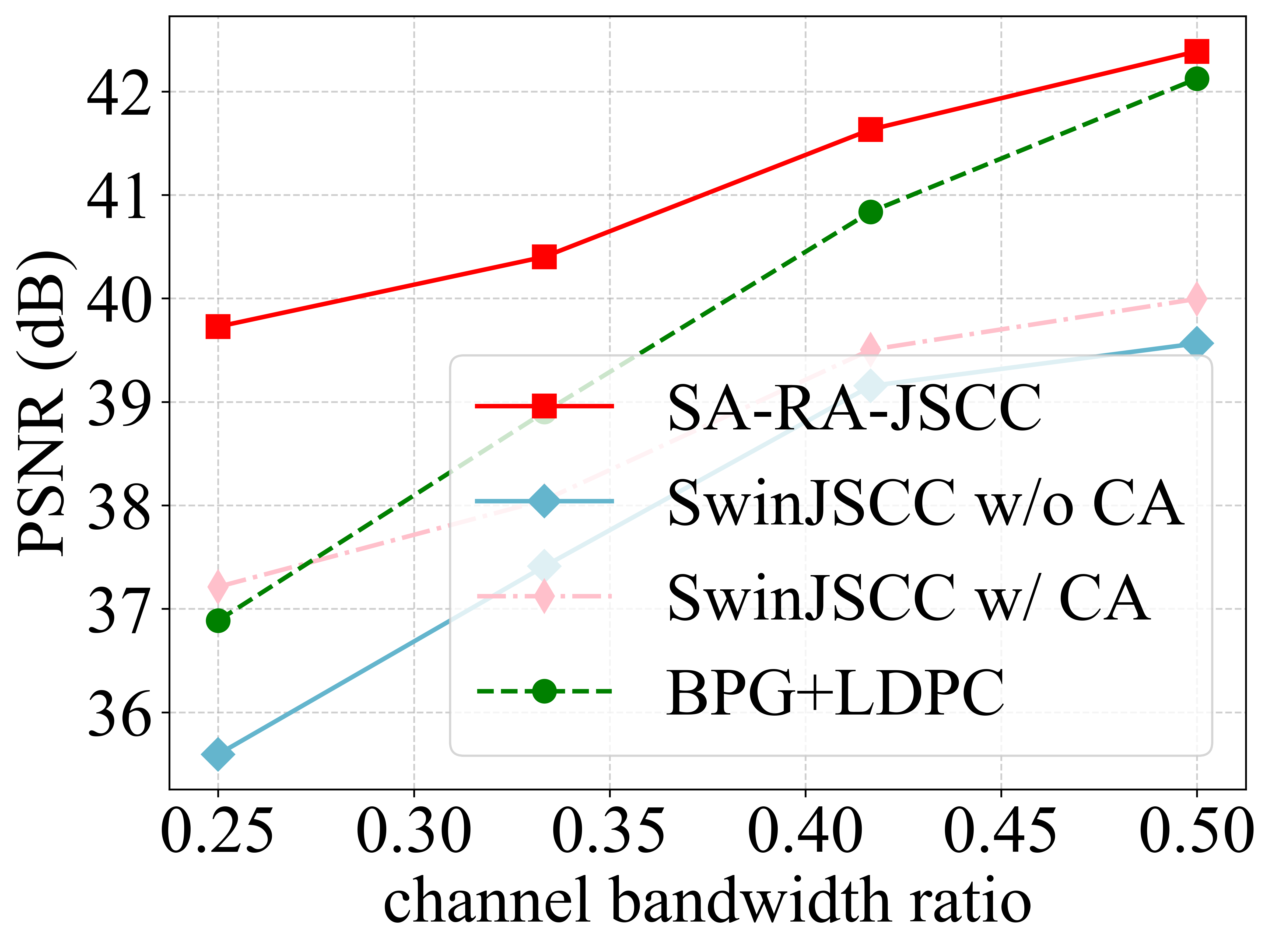}\label{fig:CLIC}}\hspace{0.5em}
  \subfloat[CLIC2020]{\includegraphics[width=0.23\textwidth]{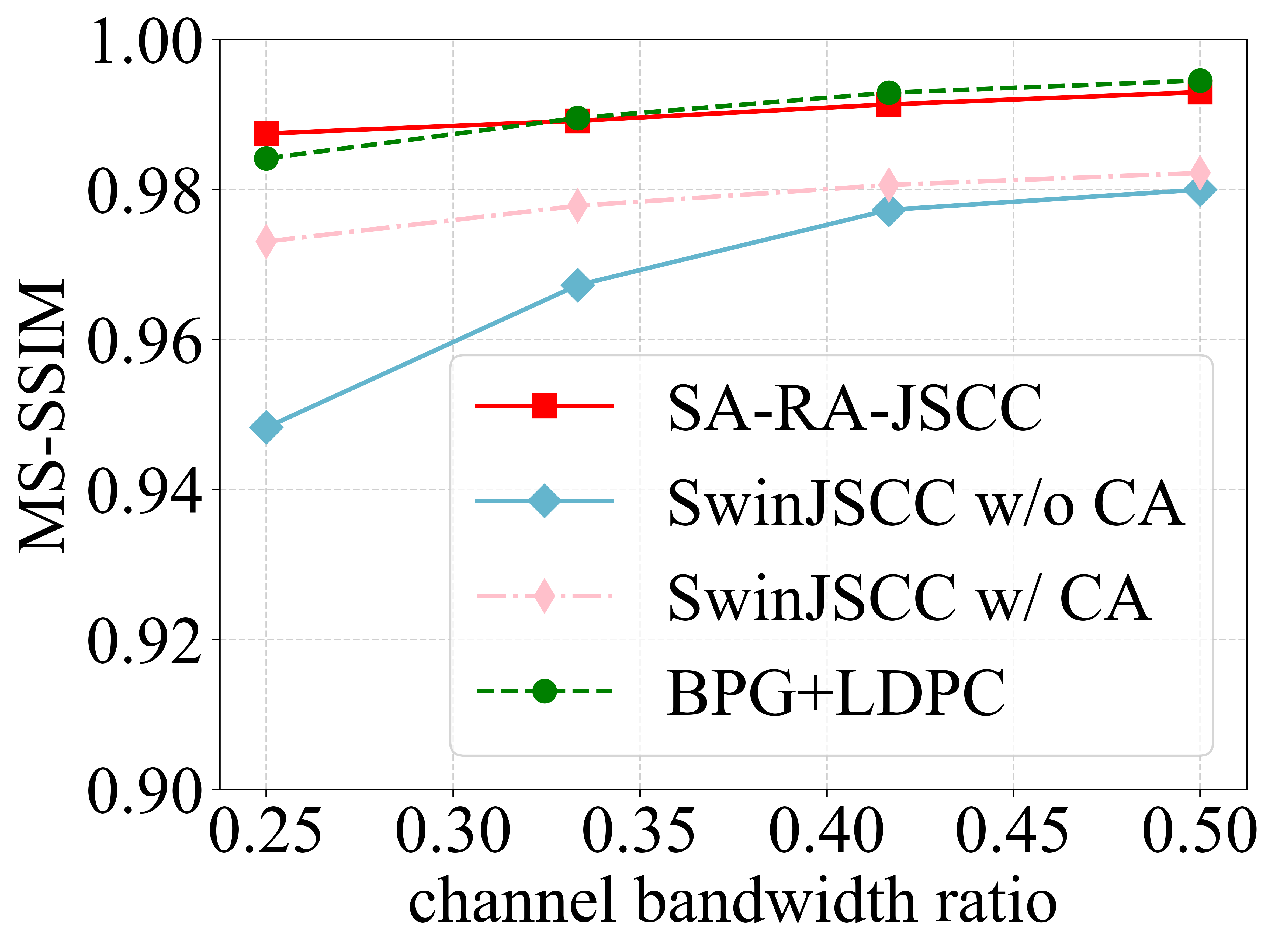}\label{fig:CLIC}}
  \caption{Comparison of different methods across various compression rates at SNR=10dB.}
  \label{Fig:4}
\end{figure}

In addition, we conducted experimental comparisons across various $cbr$ in the CLIC2020 datasets\cite{toderici2020clic}, as shown in Fig.~\ref{Fig:4}.  It can be observed that SA-RA-JSCC always achieves the best performance over the various $cbr$. Compared with SwinJSCC w/ CA and SwinJSCC w/o CA, the results indicate that simply injecting channel features in a layer-wise manner and adapting only the attention blocks are insufficient to learn more channel characteristics under different $cbr$, whereas our model yields more stable and robust reconstruction capability. Overall, these results verify that SA-RA-JSCC not only outperforms existing deep JSCC baselines, but also surpasses the conventional BPG+LDPC scheme, thereby demonstrating its clear superiority for semantic image transmission over wireless channels.

\subsection{Visualization Comparison Results}
\begin{figure*}[t]
\setlength{\abovecaptionskip}{5pt}
\centering
{\includegraphics[width=18cm,trim=2pt 2pt 2pt 2pt,clip]{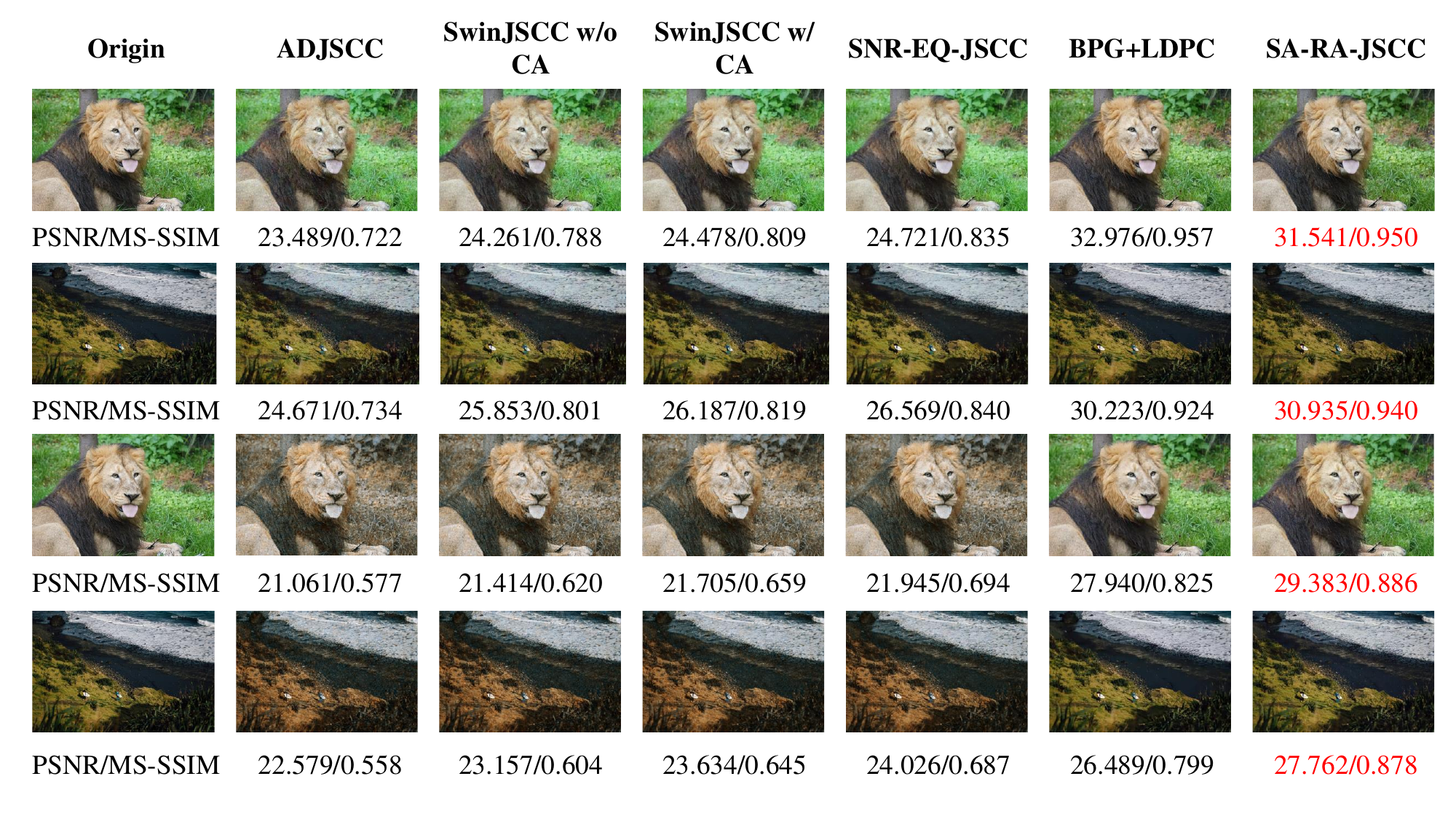}}\hspace{5pt}
\caption{Different models comparison under the SNR=10 dB of the DIV2K datasets. The first and second rows are the $r$ = 1/8, the third and fourth rows are the $r$ = 1/32.}\label{Fig:5}
\end{figure*}

To further provide an intuitive validation of the effectiveness of the proposed method, Fig.~\ref{Fig:5} presents qualitative reconstruction results on the DIV2K dataset under two settings, i.e., $r=1/8$ and $r=1/32$, at SNR = 10 dB. It can be observed that SA-RA-JSCC consistently produces reconstructions with higher visual fidelity than the competing methods under the same transmission conditions. In particular, compared with ADJSCC, SwinJSCC w/o CA, SwinJSCC w/ CA, and SNR-EQ-JSCC, the proposed method better preserves fine-grained structural information, including object contours, texture patterns, and local intensity variations. Moreover, compared with the conventional BPG+LDPC scheme, SA-RA-JSCC produces reconstructions with more stable perceptual quality in challenging cases, especially when the image contains rich local textures or complex natural structures. Although BPG+LDPC achieves competitive or even higher PSNR/MS-SSIM values in some relatively easier cases, the visual results show that SA-RA-JSCC provides more balanced semantic and perceptual preservation across different image contents.

\section{Conclusion}\label{sec:c}
This paper proposes SA-RA-JSCC, a channel-adaptive JSCC framework built upon SwinJSCC to enable stable image semantic transmission under time-varying channels constraints. Unlike prior designs that inject SNR and semantic-rate in a layer-wise and independent manner, SA-RA-JSCC maps both factors into a unified semantic vector and performs a one-shot global reweighting on the encoded features, yielding globally consistent and learnable channel adaptation. In addition, we introduced an semantic-rate module to couple channel quality variations with semantic-rate, which strengthens global coordination and adaptivity. Experiment results on multiple datasets and compression rate demonstrate that SA-RA-JSCC consistently achieves superior reconstruction quality, outperforming representative baselines in PSNR and MS-SSIM. In the future, a potential direction is to expand SA-RA-JSCC to video and voice transmission scenarios.

\begingroup
\footnotesize
\renewcommand{\baselinestretch}{1}\selectfont
\bibliographystyle{IEEEtran}
\bibliography{bibliography_ZST}
\endgroup

\end{document}